\documentclass[a4paper,10pt,oneside]{amsart}
\usepackage{geometry}
\usepackage[T1]{fontenc}
\usepackage[latin9]{inputenc}
\usepackage{amssymb}
\usepackage{amsmath}
\usepackage{amsfonts}
\usepackage{algorithm,algpseudocode}
\usepackage{xcolor}
\usepackage{graphicx}
\usepackage[colorlinks=true,linkcolor=blue,citecolor=blue,urlcolor=blue]{hyperref}
\newtheorem{proposition}{Proposition}[section]
\makeatletter
\renewcommand\paragraph{\@startsection{paragraph}{4}{\z@}{.5\linespacing\@plus.7\linespacing}{-.5em}{\normalfont\itshape}}
\makeatother
\numberwithin{equation}{section}
\DeclareMathOperator{\LSV}{LSV}
\DeclareMathOperator{\argmin}{arg{\ }min}
\newenvironment{example}[1][]
{\par\smallskip\noindent\emph{Example\if\relax\detokenize{#1}\relax\else\ (#1)\fi.}\ }
{\par\smallskip}
\newenvironment{remark}
{\par\smallskip\noindent\emph{Remark.}\ }
{\par\smallskip}
\newcommand{\dd}{\mathrm{d}}
\newcommand{\dup}{\text{Dupire}}

\makeatletter
\newenvironment{breakablealgorithm}
{%
    \pagebreak[3]
    \vspace{1em}
    \begin{center}
        \refstepcounter{algorithm}%
        \hrule height.8pt depth0pt \kern2pt%
        \renewcommand{\caption}[2][\relax]{%
                {\raggedright\textbf{\fname@algorithm~\thealgorithm} ##2\par}%
            \ifx\relax##1\relax
            \addcontentsline{loa}{algorithm}{\protect\numberline{\thealgorithm}##2}%
            \else
            \addcontentsline{loa}{algorithm}{\protect\numberline{\thealgorithm}##1}%
            \fi
            \kern2pt\hrule\kern2pt
        }
        }{%
        \kern2pt\hrule\relax
    \end{center}
    \vspace{1em}
    \pagebreak[3]
}
\makeatother
\begin{document}

\title[Local Stochastic Volatility Iterations]{Local Stochastic Volatility Iterations}

\author[C.~Bayer]{Christian Bayer}
\address{Weierstrass Institute, Berlin, Germany}
\email{christian.bayer@wias-berlin.de}

\author[D.~Belomestny]{Denis Belomestny}
\address{Duisburg-Essen University, Germany}
\email{denis.belomestny@uni-due.de}

\author[P.~P.~Hager]{Paul P. Hager}
\address{Faculty of Business, Economics and Statistics, University of Vienna, Austria}
\email{paul.peter.hager@univie.ac.at}

\author[J.~Schoenmakers]{John Schoenmakers}
\address{Weierstrass Institute, Berlin, Germany}
\email{john.schoenmakers@wias-berlin.de}


\begin{abstract}
We propose a calibration method for local stochastic volatility models that determines the leverage factor by a fixed-point iteration.
Each iteration simulates the model with the current leverage factor and approximates the conditional expectation of the stochastic variance given the spot by a single regression in time and state, with a quasi-likelihood loss for positive responses.
The resulting leverage factor is a smooth function of time and spot with relatively few parameters, which can in addition be refined directly against option quotes.
We recalibrate the leverage factor to S\&P~500 index options on the day of an interest rate announcement, keeping the parameters of a two-factor Bergomi model calibrated four weeks earlier.
After five iterations, leverage factors with $63$ and $143$ coefficients reproduce the implied volatility surface fitted to the quotes within a few basis points over the quoted strikes, the smaller one except in the left wing, and further iterations change them little.
A few regularized Gauss--Newton steps against the quotes then bring $87\%$ and $95\%$ of the liquid quotes inside their bid--ask intervals.
\end{abstract}

\maketitle

\section{Introduction}\label{sec:introduction}
Local stochastic volatility (LSV) models combine a stochastic volatility model with a local volatility factor, the leverage factor, so that the stochastic volatility governs the dynamics of the smile while the model reproduces the observed prices of vanilla options~\cite{Jex99,labordere2009calibration,RMQ07}.
For a stochastic variance process $Y$, it follows from Gy\"ongy's Markovian projection theorem~\cite{BS13,Gy86} that the LSV model $\dd S_t=S_tL(t,S_t)\sqrt{Y_t}\,\dd W_t$ has the same marginal distributions as Dupire's local volatility model~\cite{Dup94}, and hence reproduces all European call prices, if the calibration identity $L(t,x)\sqrt{\mathbb E[Y_t\mid S_t=x]}=\sigma_\dup(t,x)$ holds.
Combining this identity with the model gives
\[
    \dd S_t=S_t\,\frac{\sigma_\dup(t,S_t)}{\sqrt{\mathbb E[Y_t\mid S_t]}}\sqrt{Y_t}\,\dd W_t,
\]
which is a McKean--Vlasov equation of a singular kind.
Existence and uniqueness of its solutions are known only in special cases or under additional assumptions~\cite{AbergelTachet10,Djete22,JourdainZhou20,LackerShkolnikovZhang20,Mustapha24}, and the general case remains open.
A regularization based on reproducing kernel Hilbert spaces with propagation of chaos was proposed in~\cite{BBBS24}, and the convergence of particle schemes for regularized versions of the equation was analysed in~\cite{Mustapha24,ReisingerTsianni23,ReisingerTsianni25}.
An Euler-type time discretization of the equation is well defined, and weak error estimates for it were recently obtained in~\cite{FJWZ25}.

For stochastic volatility models with a single factor, the conditional expectation can be obtained from a two-dimensional Fokker--Planck equation~\cite{Lipton02,RMQ07}, which becomes impractical for models with several factors such as the Bergomi model~\cite{bergomi2008smileIII}.
For such models, an approximate Monte Carlo technique was proposed in~\cite{labordere2009calibration}, and the particle method~\cite{guyon2011smile,GHL14} estimates the conditional expectation at every time step by kernel regression, by averages over bins~\cite{vdSGO14}, or by least-squares regression on basis functions~\cite{Corlay16,OGH22}, and proceeds forward in time.
A leverage factor parametrized by neural networks is calibrated directly to option prices in~\cite{cuchiero2020generative}, and semimartingale optimal transport is used in~\cite{GuoLoeperWang22}.

It was suggested in~\cite{Piterbarg07} to improve an initial leverage factor $L^{(0)}$ by multiplying it with the factor $\sigma_\dup/(L^{(0)}\sqrt{\mathbb E[Y_t\mid S_t]})$, computed in the model with leverage factor $L^{(0)}$.
Repeating this step gives a fixed-point iteration, and a fixed point satisfies the calibration identity.
The idea of iterating this calibration step was suggested in~\cite[App.~B]{labordere2009ssrn}, using a Malliavin representation of the conditional expectation.
In this paper, we propose a numerical fixed-point iteration method consisting of two separate steps, the simulation of the LSV model with the current leverage factor and the approximation of the function $(t,x)\mapsto\mathbb E[Y_t\mid S_t=x]$ by regression from the simulated paths.
This separation is precisely what distinguishes our method from the particle method, which is recovered as the special case in which the regression is performed separately at every time step and the number of iterations equals the number of time steps.
The separation allows us to approximate the conditional expectation by a single regression jointly in time and state, which uses all simulated paths at once and yields a smooth leverage factor with relatively few parameters.
We use a Gamma regression with log link (see~\cite{MN89}), which keeps the approximated conditional expectation positive and makes each regression problem convex.
The method only requires samples of the stochastic variance process and therefore applies to any stochastic volatility model that can be simulated.

The relatively small number of parameters also allows us to refine the leverage factor after a few fixed-point iterations directly against the bid and ask quotes by a few regularized Gauss--Newton steps, which we call correction steps.
They largely remove the misfit between the quotes and the implied volatility surface from which $\sigma_\dup$ is computed.%

Section~\ref{sec:methodology} develops the method, from the fixed-point formulation to the regression approximation of the conditional expectation and the resulting iteration on the coefficients of a log-linear leverage factor, and estimates the pricing error of a theoretical fixed point of this iteration in Proposition~\ref{prop:pricing-error}.
Section~\ref{sec:numerics} applies it to S\&P~500 index options on 18 December 2024, the day of a decision of the Federal Open Market Committee~\cite{FOMC24}, with a two-factor Bergomi model calibrated four weeks earlier, and compares the iterates with the target surface and with the particle method.
There, a few iterations suffice for errors of a few basis points, and the correction steps then bring most liquid quotes inside their bid--ask intervals.

\section{Calibration methodology}\label{sec:methodology}
\subsection{Fixed-point formulation}

We fix a stochastic basis $(\Omega,\mathcal F,(\mathcal F_t)_{t\in[0,T]},\mathbb P)$ satisfying the usual conditions, carrying an $(\mathcal F_t)$-Brownian motion $W$ and a positive progressively measurable stochastic variance process $Y$ with $\sup_{t\in[0,T]}\mathbb E[Y_t]<\infty$.
Let $L:[0,T]\times(0,\infty)\to(0,\infty)$ be bounded and Borel measurable, and locally Lipschitz continuous in its second argument, uniformly in time.
We denote by $S=\LSV(L\vert Y,W)$ the unique positive strong solution of
\begin{equation}
    \dd S_t=S_tL(t,S_t)\sqrt{Y_t}\,\dd W_t,\qquad S_0>0,
    \label{eq:lsv}
\end{equation}
see~\cite[Th\'eor\`eme~(14.30)]{Jac79} for $\log S$,
and call $L$ the \emph{leverage factor} of the LSV model.

Our calibration target is a European call-price surface.
Under the usual no-arbitrage and regularity conditions, the Dupire construction yields a local volatility model reproducing this surface~\cite{Dup94}.
We denote the corresponding local volatility by $\sigma_\dup$ and assume that
\begin{equation}
    \dd S'_t=S'_t\sigma_\dup(t,S'_t)\,\dd B_t,\qquad S'_0=S_0,
    \label{eq:target-lv}
\end{equation}
admits a strictly positive weak solution, unique in law, which is a martingale.
The construction of the target surface and of $\sigma_\dup$ from the market quotes used in our numerical examples is described in Section~\ref{sec:numerics}.

Our goal is to choose $L$ such that $S$ and $S'$ have the same marginal distributions.
By the Markovian projection theorem~\cite{BS13,Gy86}, applied to $\log S$, this is achieved if
\begin{equation}
    L(t,x)\sqrt{\mathbb E[Y_t\mid S_t=x]}=\sigma_\dup(t,x),
    \label{eq:calibration-identity}
\end{equation}
for Lebesgue-almost every $t$ and $\mathcal L(S_t)$-almost every $x$.
The conditional expectation in~\eqref{eq:calibration-identity} depends on $L$ through the joint law of $(S,Y)$.
This suggests the following fixed-point iteration.
Starting from an initial leverage factor $L^{(0)}$, let $S^{(i)}=\LSV(L^{(i)}\vert Y,W)$ and set
\begin{equation}
    L^{(i+1)}(t,x)
    =\frac{\sigma_\dup(t,x)}{\sqrt{\mathbb E[Y_t\mid S^{(i)}_t=x]}}.
    \label{eq:fixed-point-iteration}
\end{equation}
A fixed point of~\eqref{eq:fixed-point-iteration} satisfies~\eqref{eq:calibration-identity}, cf.~\cite[App.~B]{labordere2009ssrn}.
In practice, the conditional expectation in~\eqref{eq:fixed-point-iteration} has to be approximated from simulated paths; we discuss this next.

\subsection{Conditional-expectation approximation and iterative calibration}

The fixed-point approach allows for a joint regression in time and state.
At iteration $i$, we approximate the conditional expectation
\[
    c^{(i)}(t,x):=\mathbb E[Y_t\mid S^{(i)}_t=x]
\]
within some parametrized class $\mathcal H^{(i)}$ of positive functions on $[0,T]\times (0,\infty)$.
Since $Y$ is positive, it is reasonable to consider regression losses other than least squares.
A flexible class of such losses is provided by quasi-likelihood regression, see~\cite{Wedderburn74},~\cite[Ch.~9]{MN89}.
For a positive continuous variance function $V:(0,\infty)\to(0,\infty)$, define the corresponding quasi-deviance loss
\begin{equation}
    \ell_V(y,\mu)
    :=\int_\mu^y\frac{y-u}{V(u)}\,\dd u,
    \qquad y,\mu>0.
    \label{eq:quasi-deviance}
\end{equation}
For every $V$, the conditional expectation $c^{(i)}$ is the unique minimizer of the expected loss.
Indeed, for a positive measurable function $g$ on $[0,T]\times (0,\infty)$ and $t\in[0,T]$, conditioning on $S^{(i)}_{t}$ and using~\eqref{eq:quasi-deviance} yields
\begin{equation}
    \mathbb E\bigl[\ell_V\bigl(Y_{t},g(t,S^{(i)}_{t})\bigr)\bigr]
    =\mathbb E\bigl[\ell_V\bigl(Y_{t},c^{(i)}(t,S^{(i)}_{t})\bigr)\bigr]
    +\mathbb E\bigl[\ell_V\bigl(c^{(i)}(t,S^{(i)}_{t}),g(t,S^{(i)}_{t})\bigr)\bigr],
    \label{eq:conditional-mean-loss}
\end{equation}
where the last term is nonnegative and vanishes only if $g(t,S^{(i)}_{t})=c^{(i)}(t,S^{(i)}_{t})$ almost surely.
The choice of $V$ determines how residuals at different levels of the response are weighted.

\begin{example}[Least squares]
For $V\equiv1$, \eqref{eq:quasi-deviance} gives $\ell_V(y,\mu)=\frac12(y-\mu)^2$.
\end{example}

\begin{example}[Gamma deviance]
For $V(u)=u^2$, \eqref{eq:quasi-deviance} gives $\ell_V(y,\mu)=y/\mu-\log(y/\mu)-1$.
\end{example}

Numerically, the expected loss has to be approximated from simulated paths.
Let $0=t_0<\cdots<t_n=T$ be a time grid on which the pair $(Y,W)$ is simulated.
At iteration $i$, let $\widehat S^{(i)}$ denote the approximation of $S^{(i)}$ obtained by applying a time-stepping scheme\footnote{In our numerical implementation we use an Euler--Maruyama scheme. 
Higher-order time-stepping schemes may be used in the same setup.} to~\eqref{eq:lsv} with leverage factor $L^{(i)}$ on this grid. %
Given $m$ joint samples $(Y^k,W^k)$ on the grid and the corresponding approximations $\widehat S^{(i),k}$, we replace the expected loss by the sample average over the paths, summed over the grid times with weights $w_j$, and set
\begin{equation}
    \widehat c^{(i)}
    \in \argmin_{g\in\mathcal H^{(i)}}
    \frac1m\sum_{k=1}^m\sum_{j=0}^{n}w_j\,
    \ell_V\!\left(Y^k_{t_j},g(t_j,\widehat S^{(i),k}_{t_j})\right),
    \qquad w_j\geq0.
    \label{eq:joint-regression}
\end{equation}
Any choice of a suitably trainable function class $\mathcal H^{(i)}$ in combination with the above regression loss can be used with the following iteration.
Replacing the conditional expectation in~\eqref{eq:fixed-point-iteration} by the regression approximation~\eqref{eq:joint-regression} gives the numerical fixed-point update
\begin{equation}
    L^{(i+1)}(t,x)
    =\frac{\sigma_\dup(t,x)}{\sqrt{\widehat c^{(i)}(t,x)}}.
    \label{eq:numerical-fixed-point}
\end{equation}
Each iteration consists of simulating the LSV model under the current leverage factor, approximating the conditional expectation by joint regression, and applying~\eqref{eq:numerical-fixed-point}.
The resulting procedure is summarized in Algorithm~\ref{alg:fixed-point-calibration}.

\begin{breakablealgorithm}
\caption{Fixed-point LSV calibration}\label{alg:fixed-point-calibration}
\begin{algorithmic}[1]
\Statex \textbf{Input:} target local volatility $\sigma_\dup$; initial leverage factor $L^{(0)}$; time grid $0=t_0<\cdots<t_n=T$; number of samples $m$; number of iterations $I$.
\State Generate $m$ joint samples $(Y^k,W^k)_{k=1}^m$ on the time grid.
\For{$i=0,\ldots,I-1$}
    \State Compute $\widehat S^{(i),k}$, $k=1,\ldots,m$, using the chosen time-stepping scheme with leverage factor $L^{(i)}$.
    \State Compute $\widehat c^{(i)}$ from~\eqref{eq:joint-regression}.
    \State Set $L^{(i+1)}$ according to~\eqref{eq:numerical-fixed-point}.
\EndFor
\Statex \textbf{Output:} leverage factor $L^{(I)}$.
\end{algorithmic}
\end{breakablealgorithm}

\begin{remark}
Consider the special case in which the regression is performed separately at $t_j$, $j=0,\ldots,n-1$, and $\widehat c(t,\cdot)$ is held constant on $[t_j,t_{j+1})$.
If each regression problem has a unique solution, an explicit time-stepping scheme makes the discretized fixed-point map triangular in time.
After the $i$th iteration, the leverage factor has reached its fixed-point value on the first $i$ time intervals, and Algorithm~\ref{alg:fixed-point-calibration} therefore reaches a fixed point after at most $n$ iterations.
The same fixed point can equivalently be constructed in a single forward sweep by alternating regression at $t_j$, the leverage update on $[t_j,t_{j+1})$, and propagation of the particles to $t_{j+1}$.
With a particle estimator of the conditional expectation, this gives the particle method~\cite{guyon2011smile}.
\end{remark}

\subsection{Log-linear leverage factors and the Gamma regression}\label{sec:log-linear-leverage}

In our numerical study we take $\mathcal H^{(i)}$ to be a generalized linear model with log link~\cite[Ch.~2]{MN89}. Specifically, for a vector $\psi=(\psi_1,\ldots,\psi_d)$ of real-valued basis functions on $[0,T]\times(0,\infty)$, we set
\begin{equation}
    \mathcal H^{(i)}
    =\Bigl\{(t,x)\mapsto\frac{\sigma_\dup(t,x)^2}{L^{(i)}(t,x)^2}\exp\bigl(\psi(t,x)^\top\zeta\bigr):\ \zeta\in\mathbb R^d\Bigr\}.
    \label{eq:glm-class}
\end{equation}
For this class, Algorithm~\ref{alg:fixed-point-calibration} becomes an iteration on the coefficients of a log-linear leverage factor.
Plugging the minimizer $\widehat c^{(i)}=(\sigma_\dup/L^{(i)})^2\exp(\psi^\top\zeta^{(i)})\in\mathcal H^{(i)}$ of~\eqref{eq:joint-regression} into the update step~\eqref{eq:numerical-fixed-point} yields inductively
\begin{equation}
    L^{(i+1)}%
    =L^{(0)}\exp\bigl(\psi^\top\beta^{(i+1)}\bigr),
    \qquad
    \beta^{(i+1)}=\beta^{(i)}-\tfrac12\zeta^{(i)}, \quad \beta^{(0)}=0.
    \label{eq:residual-update}
\end{equation}
In particular, the classes~\eqref{eq:glm-class} coincide, $\mathcal{H}^{(i)} = \mathcal{H}^{(0)}$ for all $i$.
We use the Gamma deviance, which renders the regression problem~\eqref{eq:joint-regression} over~\eqref{eq:glm-class} convex.
Indeed, for $y>0$ and $\mu=e^{\eta}$, $\ell_V(y,\mu)=y\,e^{-\eta}+\eta-\log y-1$ is strictly convex in $\eta$. %
Moreover, it satisfies $\ell_V(\lambda y,\lambda\mu)=\ell_V(y,\mu)$ for $\lambda>0$, which may depend on $(t,x)$.
The regression problem is therefore equivalent to minimizing over $\zeta\in\mathbb R^d$:
\begin{equation}
    \frac1m\sum_{k=1}^m\sum_{j=0}^{n}w_j\,
    \ell_V\!\left(R^{(i),k}_{t_j},\exp\bigl(\psi(t_j,\widehat S^{(i),k}_{t_j})^\top\zeta\bigr)\right),
    \qquad
    R^{(i)}_t=Y_t\,\frac{L^{(i)}(t,\widehat S^{(i)}_t)^2}{\sigma_\dup(t,\widehat S^{(i)}_t)^2}.
    \label{eq:ratio-regression}
\end{equation}

The following result estimates the pricing error of a theoretical fixed point in terms of the approximation error in the span of the basis, independently of the fixed point itself.
This may provide some further theoretical motivation for the regression formulation by considering the iteration in continuous time and with exact expectations, for $S^{(i)}=\LSV(L^{(i)}\vert Y,W)$,%
\begin{equation}
    L^{(i+1)}=\frac{\sigma_\dup}{\sqrt{\bar c^{(i)}}},
    \qquad
    \bar c^{(i)}\in\argmin_{g\in\mathcal H^{(i)}}\int_0^T\mathbb E\bigl[\ell_V\bigl(Y_t,g(t,S^{(i)}_t)\bigr)\bigr]\,\dd t.
    \label{eq:exact-iteration}
\end{equation}
Fixed-point convergence and %
convergence of %
time discretization and sampling %
are left for future work; see~\cite{BBBS24,FJWZ25,SzpruchTanTse19} for related analyses.

\begin{proposition}\label{prop:pricing-error}
Assume that $\psi_1\equiv1$ and that $\psi$ is bounded.
Suppose that the iteration~\eqref{eq:exact-iteration} admits a fixed point $L$ with finite expected loss, and let $S=\LSV(L\vert Y,W)$.
Let $f$ be a bounded payoff with maturity $T_0\leq T$ whose target value function $u_f(t,x)=\mathbb E[f(S'_{T_0})\mid S'_t=x]$, with $S'$ as in~\eqref{eq:target-lv}, belongs to $C^{1,2}([0,T_0]\times(0,\infty))$, with bounded $h_f:=\sigma_\dup^2x^2\partial_{xx}u_f=-2\partial_tu_f$, and set $h_f:=0$ on $(T_0,T]$.

Then, with  $M_{f, \psi}:= \inf_{\beta\in\mathbb R^d}\|h_f-\beta^\top\psi\|_\infty$ and $R_t=Y_tL(t,S_t)^2\sigma_\dup(t,S_t)^{-2}$ it holds that%
\begin{equation}
    \bigl|\mathbb E[f(S_{T_0})]-\mathbb E[f(S'_{T_0})]\bigr|
    \leq\frac{M_{f, \psi}}{2}\int_0^T\mathbb E\bigl[\bigl|\mathbb E[R_t\mid S_t]-1\bigr|\bigr]\,\dd t
     \le T M_{f, \psi}.
    \label{eq:price-error-bound}
\end{equation}
\end{proposition}

\begin{proof}
At the fixed point, finite expected loss and boundedness of $\psi$ justify the first-order condition $\int_0^T\mathbb E[(R_t-1)\psi(t,S_t)]\,\dd t=0$. Since $\psi_1\equiv1$, we obtain $\int_0^T\mathbb E[R_t]\,\dd t=T$ and $\int_0^T\mathbb E[|R_t-1|]\,\dd t\leq2T$. This bound and boundedness of $u_f,h_f$   justify localization and taking expectations in It\^o's formula on $[0,T_0]$ (cf.~\cite[(6.7)]{EJS98}), which together with the first-order condition yields,
\begin{equation}
    \mathbb E[f(S_{T_0})]-\mathbb E[f(S'_{T_0})] =\frac12\int_0^T\mathbb E\bigl[(R_t-1)(h_f-\beta^\top\psi)(t,S_t)\bigr]\,\dd t,
    \label{eq:price-error} \qquad \beta\in\mathbb R^d.
\end{equation}
The claim follows by conditioning on $S_t$, taking absolute values and the infimum over $\beta$.
\end{proof}

The time-stepping scheme, the initial leverage factor $L^{(0)}$, the basis functions $\psi$, the weights, the regularization and the solver used in our numerical examples are specified in Section~\ref{sec:calibration-implementations}.

\section{Calibration to SPX option data}\label{sec:numerics}

We apply the calibration method of Section~\ref{sec:methodology} to S\&P~500 index options (SPX).
We calibrate the parameters of the Bergomi model to the option quotes of 20 November 2024 and keep them fixed, while we calibrate the leverage factor to the option quotes of 18 December 2024.
On that day, the Federal Open Market Committee announced its interest rate decision~\cite{FOMC24}, and the S\&P~500 index closed $2.9\%$ lower than on the day before.
The leverage factor therefore has to reproduce a substantially changed implied volatility surface. %
As a reference method we run the particle method~\cite{guyon2011smile} on the same sample paths. 
In a second step we refine the fixed-point calibration directly against the quotes.
The implementation of the method and of this study, with all numerical settings, is publicly available at \url{https://github.com/hagerpa/locsvol}~\cite{locsvol}, together with the same study setup for other market data, a Heston backbone and the extended SSVI target of~\cite{BHP26,FFU22}.

\subsection{Market data and local volatility target}\label{sec:market-target}
The following summarizes the market data preprocessing steps and the construction of the target local volatility we use for our numerical experiments below.
For further details we refer to the implementation.

\paragraph{Quotes}
We use end-of-day bid and ask prices of SPX options from OptionMetrics IvyDB US~\cite{OptionMetrics}.
For each maturity $T$, we estimate the forward $F(T)$ and the discount factor $D(T)$ from put--call parity by least squares on mid prices, represent strikes by the log-moneyness $k=\log(K/F(T))$, divide option prices by $D(T)F(T)$, and use out-of-the-money options, with put prices converted to call prices.
We keep the quotes with log-moneyness $|k|<0.2+0.4T$, maturities $T\leq1.1$ years, and a relative bid--ask spread below $10\%$, at one expiry per calendar month and the short-dated expiries.
On 18 December, this leaves $2{,}994$ quotes at $16$ maturities between $9$ and $394$ days.

\paragraph{Weights and liquidity}
In all losses and evaluation metrics on the quotes below, every maturity and option type, put or call, receives the same total weight, shared equally among its quotes.
For the $j$th quote with maturity $T_j$ and log-moneyness $k_j$, let $\sigma_j$ be the implied volatility of its mid price and $d_j=-k_j/(\sigma_j\sqrt{T_j})+\sigma_j\sqrt{T_j}/2$.
With the standard normal distribution function $\Phi$, we denote by $v_j^{\mathrm{BS}}=\sqrt{T_j}\,\Phi'(d_j)$ its \emph{Black vega}, the derivative of the normalized Black price with respect to the volatility, so that a price difference divided by $v_j^{\mathrm{BS}}$ is, to first order, a difference in implied volatility.
We call a quote \emph{liquid} if the option traded on that day and its absolute Black delta, $\Phi(d_j)$ for a call and $\Phi(-d_j)$ for a put, is at least $0.1$\footnote{In accordance with the IvyDB volatility surface, which covers absolute deltas from $0.1$ to $0.9$~\cite{OptionMetrics}.}; on 18 December, $1{,}503$ quotes are liquid.

\paragraph{Implied volatility surface}
The fixed-point calibration requires a smooth local volatility function, which we obtain from an arbitrage-free fit of the quotes in the two steps of~\cite{gatheral2014arbitrage}.
We first fit the power-law SSVI surface, the surface form of the stochastic volatility inspired (SVI) parametrization,
\begin{equation}
    w_{\mathrm{SSVI}}(T,k)
    =\frac{\theta(T)}{2}\left(
      1+\rho\phi(T)k
      +\sqrt{\bigl(\phi(T)k+\rho\bigr)^2+1-\rho^2}
      \right),\qquad
    \phi(T)=\frac{\eta}{\theta(T)^\gamma},
    \label{eq:ssvi-surface}
\end{equation}
where $\theta(T)=w_{\mathrm{SSVI}}(T,0)$ is the at-the-money (ATM) total variance, by weighted least squares in implied volatility under the sufficient conditions of~\cite{gatheral2014arbitrage} for the absence of static arbitrage, with $\gamma\leq1/2$.
Starting from this fit, we refine every maturity slice by the raw SVI parametrization $w(T,k)=a_T+b_T\bigl(\rho_T(k-m_T)+\sqrt{(k-m_T)^2+\sigma_T^2}\bigr)$, again by weighted least squares in implied volatility, subject to butterfly and calendar constraints on a grid of log-moneyness values.
Between maturities, we interpolate as in~\cite{gatheral2014arbitrage}, linearly in the square root of the ATM total variance, and beyond the last maturity the total variance increases linearly.

\paragraph{Local volatility target}
We denote by $\widehat{\mathcal V}(T)=2\int e^{-k}\,\mathrm{BS}\bigl(k,w(T,k)\bigr)\,\dd k$ the total variance of the log-contract, that is, $T$ times the fair variance-swap rate~\cite{DemeterfiEtAl99},~\cite[Ch.~11]{Gatheral06}, where $\mathrm{BS}(k,w)$ denotes the normalized Black price of the out-of-the-money option with log-moneyness $k$ and total variance $w$.
Below the first maturity $T_1=9$ days, we use the extension
\[
    w(t,k)=\frac{t}{T_1}\Bigl(1+c\Bigl(1-\frac{t}{T_1}\Bigr)\Bigr)\,
    w\biggl(T_1,\,k\sqrt{\frac{T_1+t_0}{t+t_0}}\biggr),
\]
where $t_0$ and $c$ are chosen such that the forward variance $\partial_T\widehat{\mathcal V}$ is continuous at $T_1$.
The target local volatility $\sigma_\dup$ follows from the refined SVI surface by the Dupire formula in terms of total variance~\cite[Ch.~1]{Gatheral06}.

\subsection{Bergomi stochastic-volatility specification}\label{sec:bergomi-specification}

The stochastic variance factor follows the two-factor Bergomi model~\cite{bergomi2005smileII,bergomi2008smileIII}.
Let $Z^1$ and $Z^2$ be Brownian motions with $\langle Z^1,Z^2\rangle_t=\rho_Z\,t$ and $\langle W,Z^a\rangle_t=\rho_a\,t$, $a=1,2$, and define $Y_t=\exp\bigl(2\nu G_t-2\nu^2\,\mathbb E[G_t^2]\bigr)$, with
\begin{equation}
\begin{aligned}
    X_t^a&=\int_0^t e^{-\kappa_a(t-s)}\,\dd Z_s^a,\quad (a=1,2),
    \qquad
    G_t=\frac{(1-\vartheta)X_t^1+\vartheta X_t^2}{\sqrt{(1-\vartheta)^2+\vartheta^2+2\rho_Z\vartheta(1-\vartheta)}},
\end{aligned}
    \label{eq:bergomi-factor}
\end{equation}
Then $\mathbb E[Y_t]=1$.
The factors $X^1$ and $X^2$ are simulated exactly on the time grid, jointly with the increments of $W$.

We fix $\kappa_1=4$, $\kappa_2=0.12$ and $\rho_Z=0.3$, and calibrate $(\nu,\vartheta,\rho_1,\rho_2)$ to the quotes of 20 November with maturities beyond $18$ days in the pure stochastic volatility model with variance $\xi\,Y$, where the forward variance curve comes from the SVI surface of that day via
\[
    \xi(t)=\bigl(\widehat{\mathcal V}(t_{j+1})-\widehat{\mathcal V}(t_j)\bigr)/(t_{j+1}-t_j),\qquad t_j\leq t<t_{j+1}.
\]
We minimize the vega-weighted squared differences between model and mid prices by the same Gauss--Newton procedure as described in Section~\ref{sec:calibration-implementations}, and obtain
\[
    (\nu,\vartheta,\rho_1,\rho_2)=(2.3192,\;0.1685,\;-0.7541,\;-0.2790).
\]
When we use the model as the \emph{backbone} of the LSV model~\eqref{eq:lsv} on 18 December, we drop the forward variance, use the factor $Y$ alone and initialize the iteration with $L^{(0)}$ as specified below.
Equivalently, one could use $\xi\,Y$ as a backbone and initialize with $L^{(0)}/\sqrt{\xi}$.

\subsection{Calibration implementations}\label{sec:calibration-implementations}

\paragraph{Leverage parametrization}
We use the log-linear leverage factors~\eqref{eq:residual-update} of Section~\ref{sec:log-linear-leverage}.
We start the fixed-point iteration from the Dupire local volatility read within the quantiles of the SSVI fit, $L^{(0)}(t,e^{x})=\sigma_\dup\bigl(t,e^{\bar x}\bigr)$ with $\bar x=\max(x_\varepsilon(t),\min(x,x_{1-\varepsilon}(t)))$, where $x_\varepsilon(t)$ and $x_{1-\varepsilon}(t)$ are the $\varepsilon$- and $(1-\varepsilon)$-quantiles of $\log S_t$ under the SSVI surface~\eqref{eq:ssvi-surface}, and use the basis functions
\begin{equation}
    \psi(t,e^{x})=\Bigl(C_p\bigl(\tau(t)\bigr)\,C_q\bigl(\widetilde u(t,x)\bigr)\Bigr)_{0\leq p\leq P,\ 0\leq q\leq Q}\in\mathbb R^d,\qquad d=(P+1)(Q+1),
    \label{eq:leverage-parametrization}
\end{equation}
where $C_p$ denotes the Chebyshev polynomial of degree $p$ and $\tau$, $\widetilde u$ are the coordinates~\eqref{eq:target-adapted-coordinates} below.
We use the time--state degrees $(P,Q)=(6,8)$, with $63$ coefficients, and $(P,Q)=(10,12)$, with $143$ coefficients.

To adapt the basis to the target surface, we use the SSVI parameters of~\eqref{eq:ssvi-surface}\footnote{We regard the fit of the four SSVI parameters $\theta,\rho,\eta,\gamma$ as a preliminary calibration of our regression basis.}, extend $\theta$ linearly beyond the last maturity, hold it constant near $t=0$, which keeps the coordinates finite there, and define
\begin{equation}
    u(t,x)=\frac{2}{\pi}\arctan\!\left(\frac{\eta\,x}{\theta(t)^{\gamma}}\right),
    \qquad
    \tau(t)=2\,\frac{\chi(t)-\chi(0)}{\chi(T)-\chi(0)}-1,
    \qquad
    \chi(t)=\frac{\theta(t)^{\gamma}}{\eta}.
    \label{eq:target-adapted-coordinates}
\end{equation}
Since $\chi=1/\phi$, the coordinate $u$ measures the log price $x$ in units of the width of the target smile at time $t$, and $\tau$ uses the same scale as a clock.
Only a few paths reach the states beyond the quantiles, where the regression does not determine a polynomial in $u$.
We therefore use the coordinate $\widetilde u(t,x)$, which is $u(t,x)$ clipped to $[u(t,x_\varepsilon(t)),u(t,x_{1-\varepsilon}(t))]$ with $\varepsilon=10^{-4}$ and mapped affinely onto $[-1,1]$, so that $\exp(\psi^\top\beta)$ is constant in the state beyond the quantiles.

\paragraph{Regression}
We add the ridge penalty $2.5\cdot10^{-7}\lVert\zeta\rVert^2/2$ to the regression~\eqref{eq:ratio-regression}, which makes it strictly convex with a unique minimizer, and solve it by L-BFGS~\cite[Sec.~7.2]{NW06}.

\paragraph{Sample paths and time grid}
We discretize~\eqref{eq:lsv} by the Euler--Maruyama scheme for $\log S$ on a time grid that contains the points $(90/365)(l/800)^2$, $l=0,\ldots,800$, and then points with uniform spacing $T/440$ up to $T=1.1$ years. %
The regression uses $m=2^{15}$ sample paths generated from randomized Sobol points with a Brownian bridge construction and antithetic variates~\cite[Secs.~3.1, 4.2, 5.2.3 and 5.4]{Glasserman04},~\cite{LEcuyer09}, which are kept fixed in all iterations.
In~\eqref{eq:joint-regression}, we use four times the trapezoidal weights of the time grid, $w_j=2(t_{j+1}-t_{j-1})/T$ with $t_{-1}:=t_0$ and $t_{n+1}:=t_n$, at a fourth of the time points, evenly spread along each path, and set all other weights to zero.

\paragraph{Particle method}
As a reference, we use the particle method~\cite{guyon2011smile} with the same Bergomi model, target local volatility, time grid and sample paths, and the settings of~\cite[Sec.~11.6]{GHL14}.
At each grid time, the conditional expectation is the Nadaraya--Watson estimate with the quartic kernel $K(z)=\frac{15}{16}(1-z^2)^2\mathbf 1_{\{|z|\leq1\}}$ and bandwidth $1.5\,(\widehat{\mathcal V}(t)\max(t,1/4))^{1/2}\,t^{-1/2}\,m^{-1/5}$ at $\lceil\max(30\sqrt t,15)\rceil$ points between the $0.001$ and $0.999$ quantiles of the refined SVI target distribution, and the resulting leverage factor is interpolated by a cubic spline, extrapolated flat and used in absolute value.

\paragraph{Option-price correction}
The leverage factor~\eqref{eq:residual-update} can also be calibrated directly to the quotes by Monte Carlo, as~\cite{cuchiero2020generative} do for neural networks without the calibration identity.
Since our parametrization has few coefficients, we can efficiently use Gauss--Newton steps~\cite[Ch.~10]{NW06} with the full pathwise Jacobian~\cite{BG96},~\cite[Sec.~7.2]{Glasserman04} of the Monte Carlo prices, and we study this refinement started from different iterates of the fixed-point iteration.
For prices $p=(p_1,\ldots,p_N)^\top$, intervals $[P_j^{\mathrm b},P_j^{\mathrm a}]$ and the quote weights $\omega_j$, we use the bid--ask loss
\begin{equation}
    \mathcal E_{\mathrm{BA}}(p)^2
    =\sum_{j=1}^N\omega_j\left(\frac{\operatorname{dist}\bigl(p_j,[P_j^{\mathrm b},P_j^{\mathrm a}]\bigr)}{v_j^{\mathrm{BS}}}\right)^2
    \label{eq:bid-ask-loss}
\end{equation}
For a liquid quote, $[P_j^{\mathrm b},P_j^{\mathrm a}]$ is its bid--ask interval; for every other quote, it is extended to contain the price at the start of the correction, so that the quote may approach its interval at no cost but may not move further away from it.
The steps are restricted to a trust region~\cite{CGT00},~\cite[Ch.~4]{NW06}, measured by the mean-square change of $\log L$ along the simulated paths, since the linearized prices are accurate only near the current coefficients.
The corrected leverage factor is also kept within a factor $e^{\pm0.1}$ of the starting iterate over all steps.
We start after $0$, $3$, $5$ and $8$ fixed-point iterations and apply four steps.

\subsection{Numerical results}\label{sec:numerical-results}

To evaluate the calibrated models, we simulate $320$ independent replications of $2^{16}$ pseudo-random sample paths, independent of all paths used for calibration and common to all models, and average the prices before computing implied volatilities.
For the comparison with the target, we denote by $k_{\min}(T)<0<k_{\max}(T)$ the smallest and the largest log-moneyness of the quotes with maturity $T$ and use $81$ equally spaced log strikes per maturity between them.
We assign a log strike $k$ to the \emph{left wing} if $k\leq0.4\,k_{\min}(T)$, to the \emph{right wing} if $k\geq0.4\,k_{\max}(T)$, and to the \emph{central region} otherwise, and within each region every maturity receives the same weight.
We report the root mean square difference between model and target implied volatilities in volatility basis points (bp); their Monte Carlo standard error is about $3.2$~bp in the left wing and below $1$~bp elsewhere.
The fit to the market is measured by~\eqref{eq:bid-ask-loss} with the bid--ask intervals, in bp, over all quotes and over the liquid quotes.

\begin{figure}[htbp]
    \centering
    \includegraphics[width=0.9\textwidth,trim=0 8.5 0 6.7,clip]{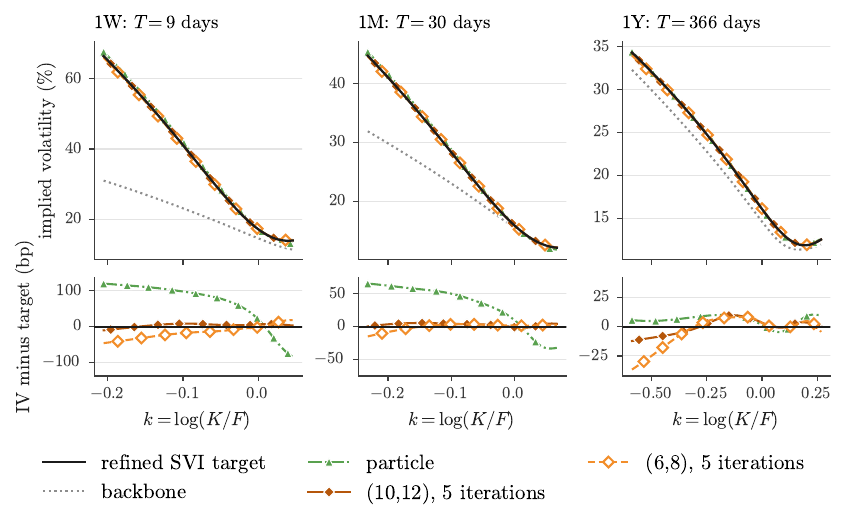}
    \caption{Implied volatility smiles of SPX options on 18 December 2024 at $9$, $30$ and $366$ days.
    Top: the refined SVI target, the Bergomi model as calibrated on 20 November, labelled backbone, the particle method, and the joint regression with degrees $(10,12)$ and $(6,8)$ after five fixed-point iterations.
    Bottom: differences to the target in volatility basis points.}
    \label{fig:fixed-point-smiles}
\end{figure}

\paragraph{Fixed-point calibration}
In the smiles at $9$, $30$ and $366$ days in Figure~\ref{fig:fixed-point-smiles}, the Bergomi model as calibrated on 20 November, labelled backbone, deviates substantially from the target of 18 December, in the level as well as in the shape of the smile.
The particle method deviates from the target mainly at short maturities; its accuracy depends on the number of particles and on the bandwidth, which we do not tune.

\begin{figure}[htbp]
    \centering
    \includegraphics[width=0.9\textwidth,trim=0 7.6 0 6.3,clip]{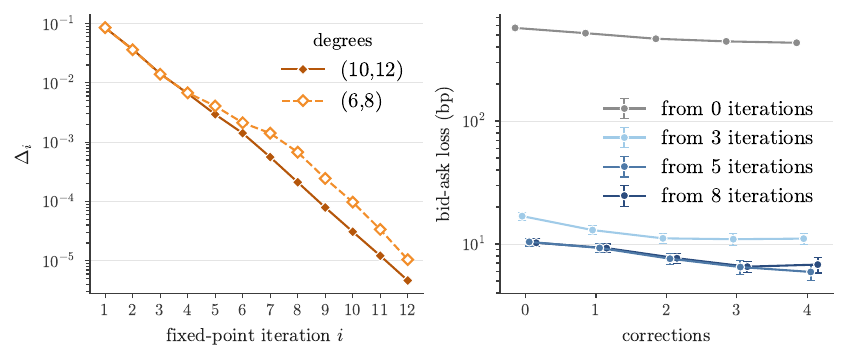}
    \caption{Left: root mean square change $\Delta_i$ of the leverage factor along the simulated paths for the degrees $(10,12)$ and $(6,8)$.
    Right: bid--ask loss over all quotes along the correction steps for the degrees $(6,8)$, started after $0$, $3$, $5$ and $8$ iterations, $\pm2$ standard errors from $20$ independently randomized sets of $2^{16}$ Sobol points.
    Both axes are logarithmic.}
    \label{fig:fixed-point-convergence}
\end{figure}

The left panel of Figure~\ref{fig:fixed-point-convergence} shows the root mean square change of the leverage factor along the simulated paths in every iteration, \[\Delta_i^2=\sum_{j=0}^{n}\frac{t_{j+1}-t_{j-1}}{2T}\,\frac1m\sum_{k=1}^m\bigl(L^{(i)}-L^{(i-1)}\bigr)^2(t_j,\widehat S^{(i),k}_{t_j}).\]
Its decrease suggests a geometric decay, by a factor of roughly one half per iteration, to $10^{-5}$ or below after $12$ iterations for both degrees.
After five iterations, the errors in Table~\ref{tab:target-errors} are below $5$~bp in all three regions with degrees $(10,12)$, and with degrees $(6,8)$ they are about $5$ and $6$~bp in the central region and the right wing but $17$~bp in the left wing.
The errors change by at most about two basis points in further iterations.

\begin{table}[htbp]
    \centering
    \caption{Root mean square error of the implied volatility against the refined SVI target in volatility basis points after five fixed-point iterations, over the left wing $k\leq0.4\,k_{\min}(T)$, the central region and the right wing $k\geq0.4\,k_{\max}(T)$, with the particle method as a reference, as point estimates $\pm2$ Monte Carlo standard errors.}
    \label{tab:target-errors}
    \begin{tabular}{lccc}
        \hline
        Method & Left & Central & Right \\
        \hline
        Joint regression $(10,12)$, 5 iterations & $4.58\pm2.13$ & $4.72\pm0.94$ & $3.59\pm0.44$ \\
        Joint regression $(6,8)$, 5 iterations & $16.91\pm4.16$ & $4.86\pm0.73$ & $6.20\pm0.35$ \\
        Particle method & $46.47\pm3.97$ & $26.17\pm1.01$ & $26.91\pm0.32$ \\
        \hline
    \end{tabular}
\end{table}

\paragraph{Option-price correction}
Figure~\ref{fig:fixed-point-convergence} and Table~\ref{tab:corrections} show the bid--ask losses along and after the four correction steps.
Started from $L^{(0)}$, the bounded steps reduce the loss over all quotes only from $573$ to $434$~bp for $(6,8)$ and to $433$~bp for $(10,12)$, as perhaps expected for a local refinement.
Started after five iterations, they reduce the loss over the liquid quotes from $6.25$ to $0.81$~bp for $(6,8)$ and from $5.49$ to $0.30$~bp for $(10,12)$, and the share of liquid quotes inside their bid--ask intervals from $45\%$ to $87\%$ and from $50\%$ to $95\%$, respectively.
Before the correction, both regressions fit the liquid quotes about as well as the target itself, which does not lie inside the bid--ask intervals.
The correction largely removes this misfit with the same $63$ or $143$ coefficients, moving the models away from the target to errors of $10$ to $12$~bp for $(6,8)$ and $9$ to $13$~bp for $(10,12)$.

\begin{table}[htbp]
    \centering
    \caption{Bid--ask loss~\eqref{eq:bid-ask-loss} in volatility basis points and share of liquid quotes inside their bid--ask interval after four correction steps, as point estimates $\pm2$ Monte Carlo standard errors.}
    \label{tab:corrections}
    \begin{tabular}{llccc}
        \hline
        Degrees & Correction started after & All quotes & Liquid quotes & Liquid inside \\
        \hline
        $(10,12)$ & $0$ iterations & $433.29\pm1.50$ & $258.78\pm0.40$ & $4\%$ \\
         & $3$ iterations & $7.20\pm1.15$ & $0.52\pm0.13$ & $87\%$ \\
         & $5$ iterations & $3.46\pm0.74$ & $0.30\pm0.29$ & $95\%$ \\
         & $8$ iterations & $3.37\pm0.70$ & $0.29\pm0.28$ & $95\%$ \\
        \hline
        $(6,8)$ & $0$ iterations & $434.26\pm1.50$ & $259.16\pm0.40$ & $3\%$ \\
         & $3$ iterations & $11.16\pm1.32$ & $0.85\pm0.14$ & $81\%$ \\
         & $5$ iterations & $5.99\pm0.97$ & $0.81\pm0.45$ & $87\%$ \\
         & $8$ iterations & $6.68\pm1.05$ & $0.96\pm0.47$ & $84\%$ \\
        \hline
        \multicolumn{2}{l}{Refined SVI target} & $7.37$ & $6.40$ & $42\%$ \\
        \hline
    \end{tabular}
\end{table}

In summary, after five fixed-point iterations a leverage factor with $63$ coefficients reproduces the target surface within a few basis points at the money and in the right wing, one with $143$ coefficients also in the left wing, and four correction steps of these coefficients bring $87\%$ and $95\%$ of the liquid quotes inside their bid--ask intervals.
The repository contains the same study for four further days with market events, with the Bergomi and a Heston backbone, and a calibration to the extended SSVI surface of~\cite{FFU22} for SPX options of 23 June 2020 with maturities up to three years, the target of~\cite{BHP26}, with the reference models of that paper.

\section*{Acknowledgments}
We thank the Vienna Data Center of the University of Vienna for access to the OptionMetrics data.
Wharton Research Data Services (WRDS) was used in preparing this paper.
This service and the data available thereon constitute valuable intellectual property and trade secrets of WRDS and/or its third-party suppliers.

\end{document}